\documentclass[useAMS,usenatbib]{mn2e}

\usepackage{amssymb,amsmath,amsfonts,amsbsy,graphicx}

\def\mathbi#1{\textbf{\em #1}}

\newcommand{\fnl}{f_{\rm NL}}
\newcommand{\gnl}{g_{\rm NL}}
\newcommand{\tnl}{\tau_{\rm NL}}

\newcommand{\calF}{{\cal F}}
\newcommand{\calG}{{\cal G}}
\newcommand{\calM}{{\cal M}}

\newcommand{\calT}{{\cal T}}

\title[Scale dependent bias from primordial non-Gaussianity with trispectrum]
{Scale dependent bias from primordial non-Gaussianity with trispectrum}
\author[J.~O.~Gong and S.~Yokoyama]{Jinn-Ouk Gong$^{1}$
and Shuichiro Yokoyama$^{2}$\\
$^{1}$Theory Division, CERN, CH-1211 Gen\`eve 23, Switzerland
\\
$^{2}$Department of Physics and Astrophysics, Nagoya University,
Aichi, 464-8602, Japan
}

\begin{document}

\date{Accepted. Received; in original form}

\pagerange{\pageref{firstpage}--\pageref{lastpage}} \pubyear{2011}

\maketitle

\label{firstpage}

\begin{abstract}

We study the scale dependent bias of the halo power spectrum arising from primordial non-Gaussianity.
We present an analytic result of the halo bias including up to the trispectrum contributions.
We find the scale dependent bias opens a new possibility of probing the relation between the non-linearity parameters $\fnl$ and $\tnl$.

\end{abstract}

\begin{keywords}
cosmology: theory -- large-scale structure of the Universe
\end{keywords}

\section{Introduction}

Inflation~\citep{1981PhRvD..23..347G,1981MNRAS.195..467S,1982PhLB..108..389L,1982PhRvL..48.1220A} is currently 
regarded as the leading candidate of solving the cosmological problems of the hot big bang 
universe and of laying down the otherwise finely tuned initial conditions. Furthermore,
quantum fluctuations of the inflaton fields are stretched to cosmological scales during inflation
and become the seed of subsequent structure formation after inflation.
An important prediction of the standard single field slow-roll inflation scenario is that
the initial perturbations are Gaussian random fields.
On the other hand, there are a lot of theoretical models of generating the primordial fluctuations 
during inflation which deviate from the Gaussian statistics. 
For recent reviews, see e.g.~\citet{2010CQGra..27l0301S}.
Although the precise measurements of the cosmic microwave background (CMB) anisotropies over the last decade~\citep{2011ApJS..192...18K} strongly suggest that the initial perturbations generated during 
inflation follow nearly perfect Gaussianity,
there still remains the possibility that we will detect small deviation
by non-zero higher order correlation functions.
Thus, detection or null-detection of non-Gaussianity of primordial perturbations in future 
experiments will play a crucial role in discriminating inflationary models~\citep{2009astro2010S.158K}.

Besides the CMB, ongoing and planned survey of large scale structure (LSS) will provide 
another powerful probe of 
constraining non-Gaussianity as competitive as the CMB~\citep{2009astro2010S.158K,2010AdAst2010E..64V}.
In particular, it is known that the local type non-Gaussianity 
$\fnl\phi^2$, 
with $\phi$ being the dominant Gaussian component of the Bardeen potential $\Phi$, induces a scale dependent bias~\citep{2008PhRvD..77l3514D,2008ApJ...677L..77M}.
This is an interesting way of probing primordial non-Gaussianity, since future surveys will 
provide a large enough sample of galaxies over a huge volume.

With the promised precise data from LSS surveys,
we are encouraged to go beyond the leading non-Gaussianity.
$\fnl\phi^2$ gives the leading order 3-point correlation function, or the bispectrum.
To describe the 4-point correlation function, or the trispectrum,
unlike the bispectrum we need to specify two parameters, $\gnl$ and $\tnl$~\citep{2006PhRvD..73b1301B,2006PhRvD..74l3519B}.
While $\gnl$ is the local cubic expansion parameter of $\Phi$, $\tnl$ may or may not be related to $\fnl$:
if primordial non-Gaussianity is sourced by a single origin, it can be solely written in terms of $\fnl$~\citep{2006PhRvD..73b1301B}.
But in general there is no universal relation between $\fnl$ and $\tnl$~\citep{2008PhRvD..77b3505S,2010JCAP...12..030S,2011arXiv1101.3636S}.
Thus, it is very interesting and potentially important observationally to study the consequences of generic $\tnl$ which is independent of $\fnl$, such as the halo mass function~\citep{2011arXiv1103.2586Y}.
Recently two interesting articles appear with similar viewpoints to ours. In~\citet{2011arXiv1105.3628D}, general expressions of bias with corrections from higher order correlation functions were given. \citet{2011arXiv1106.0503S} considered the halo bias arising from $\gnl$. But $\tnl$ and its relation to $\fnl$ were not examined in both articles.

In this short article, we study analytically the halo bias in the presence of primordial local type non-Gaussianity
with non-vanishing trispectrum characterized by two parameters $\gnl$ and $\tnl$. 
Very interestingly, we find that the scale dependent bias arising from trispectrum enables us to
test the relation between $\gnl$ and $\tnl$.
This article is outlined as follows. In Section~\ref{sec:nongaussian}, we recall the primordial
non-Gaussianity up to trispectrum. In Section~\ref{sec:halopower} we present the power spectrum of
haloes, and then derive an analytic result of halo bias in Section~\ref{sec:dependentbias}.
In Section~\ref{sec:conclusion}, we conclude.

\section{Primordial non-Gaussianity}
\label{sec:nongaussian}

In the presence of the local type non-Gaussianity, we may expand the Bardeen potential $\Phi$ up to the third order as a local function,
\begin{equation}\label{Phi}
\Phi(\mathbi{x}) = \phi(\mathbi{x}) + \fnl \left[ \phi^2(\mathbi{x}) - \left\langle\phi^2\right\rangle \right] + \gnl \phi^3(\mathbi{x}) \, .
\end{equation}
Then, with the linear power spectrum of $\Phi$ defined by
\begin{equation}
\langle \phi(\mathbi{k}_1)\phi(\mathbi{k}_2)\rangle =
 (2\pi)^3 \delta (\mathbi{k}_1 + \mathbi{k}_2) P_\phi(k_1) \, ,
\end{equation}
at leading order we can find the bispectrum and trispectrum of $\Phi$ straightforwardly as
\begin{align}
& B_\Phi(\mathbi{k}_1,\mathbi{k}_2,\mathbi{k}_3) = 2\fnl \left[ P_\phi(k_1)P_\phi(k_2) + \text{(2 cyclic)} \right] \, ,
\\
& T_\Phi(\mathbi{k}_1,\mathbi{k}_2,\mathbi{k}_3,\mathbi{k}_4) = 6\gnl \left[ P_\phi(k_1)P_\phi(k_2)P_\phi(k_3) + \text{(3 cyclic)} \right]
\nonumber\\
& + 2\fnl^2 \left\{ P_\phi(k_1)P_\phi(k_2) \bigl[ P_\phi(k_{13})
+ P_\phi(k_{14}) \bigr] +  \text{(11 cyclic)} \right\} \, ,
\end{align}
with $\mathbi{k}_{ij} \equiv \mathbi{k}_i + \mathbi{k}_j$.
While the cubic order term in (\ref{Phi}) does not appear in the leading order bispectrum,
it does generate the primordial trispectrum.
Here, we can generalize the second term of the trispectrum by replacing $\fnl^2$ with a new parameter $\tnl$, which may or may not be related to $\fnl$, as
\begin{align}
& T_\Phi(\mathbi{k}_1,\mathbi{k}_2,\mathbi{k}_3,\mathbi{k}_4) = 6\gnl \left[ P_\phi(k_1)P_\phi(k_2)P_\phi(k_3) + \text{(3 cyclic)} \right]
\nonumber\\
& + \frac{25}{18}\tnl \left\{ P_\phi(k_1)P_\phi(k_2) \bigl[ P_\phi(k_{13}) + P_\phi(k_{14}) \bigr] +  \text{(11 cyclic)} \right\} \, .
\end{align}
Note that the coefficient of $\tnl$ reflects the definition of $\tnl$ introduced in~\citet{2006PhRvD..73b1301B}. This generalization of the non-linearity parameter $\tau_{\rm NL}$
gives rise to the local type inequality given by~\citep{2008PhRvD..77b3505S}
\begin{equation}\label{ineq}
\tnl \geq \left( \frac{6}{5}\fnl \right)^2 \, .
\end{equation}

\section{Power spectrum of haloes}
\label{sec:halopower}

In this section, we derive analytically the power spectrum of haloes arising from the local type non-Gaussianity (\ref{Phi}).
We will use the functional integration approach~\citep{1984ApJ...285L...1P,1986ApJ...310...19G,1986ApJ...310L..21M} for the correlation functions of peaks of the density field.

We can relate the linear density field $\delta_R(k)$ smoothed over radius $R$ to $\Phi(k)$ via the Poisson equation,
\begin{equation}
\delta_R(k) = \frac{2}{3}\frac{k^2T(k)}{H_0^2\Omega_{m0}}W_R(k)\Phi(k) \equiv \calM_R(k)\Phi(k) \, ,
\end{equation}
where $T(k)$ is the matter transfer function, $H_0$ is the present Hubble parameter,
$\Omega_{m0}$ is the present matter density parameter and $W_R(k)$ is the Fourier transform of the window function.
Then, we can write the bispectrum and the trispectrum of $\delta_R(k)$ as
\begin{align}
\label{bihalo}
& B_R(\mathbi{k}_1,\mathbi{k}_2,\mathbi{k}_3) = 2\fnl \prod_{i=1}^3 \calM_R(k_i) \left[ P_\phi(k_1)P_\phi(k_2) + \text{(2 cyclic)} \right] \, ,
\\
\label{trihalo}
& T_R(\mathbi{k}_1,\mathbi{k}_2,\mathbi{k}_3,\mathbi{k}_4) =
\nonumber\\
& 6\gnl \prod_{i=1}^4 \calM_R(k_i) \left[ P_\phi(k_1)P_\phi(k_2)P_\phi(k_3) + \text{(3 cyclic)} \right]
\nonumber\\
& + \frac{25}{18}\tnl \prod_{i=1}^4 \calM_R(k_i) \left\{ P_\phi(k_1)P_\phi(k_2) \left[ P_\phi(k_{13}) + P_\phi(k_{14}) \right] \right.
\nonumber\\
& \left. \hspace{3cm} + \text{(11 cyclic)} \right\} \, .
\end{align}

Employing the functional integration approach for distributions of the haloes above a high threshold, the two-point correlation function of haloes with generic non-Gaussian density field can be written as~\citep{1986ApJ...310...19G,1986ApJ...310L..21M,2008ApJ...677L..77M}
\begin{equation}\label{2pointcorr}
\xi_h(\mathbi{x}_1,\mathbi{x}_2) = \exp \left[ \sum_{n=2}^\infty \sum_{m=1}^{n-1} \frac{\nu^nw_m^{(n)}}{m!(n-m)!} \right] - 1 \, .
\end{equation}
Here, $\nu \equiv \delta_c/\sigma_R$ with $\delta_c$ being the critical density and $\sigma_R$ being the variance of the density field $\delta_R$, and the coefficient $w_m^{(n)}$ is given by
\begin{equation}\label{coeff_wmn}
w_m^{(n)} = \left\{
\begin{split}
 & \frac{\xi_R^{(2)}(r)}{\sigma_R^2} & (n=2\,,m=1)
 \\
 & 0 & (n=2\,,m=0 \, , 2)
 \\
 & \frac{\xi_{R,m}^{(n)}}{\sigma_R^n} & (n>2)
\end{split}
\right. \, ,
\end{equation}
where $\mathbi{r} \equiv \mathbi{x}_1-\mathbi{x}_2$, $\xi_R^{(n)}$ is the connected $n$-point correlation function of $\delta_R$ and
\begin{equation}
\xi_{R,m}^{(n)} \equiv \xi_R^{(n)}(\underbrace{\mathbi{x}_1,\mathbi{x}_1,\cdots\mathbi{x}_1}_{\text{total $m$}},
\underbrace{\mathbi{x}_2,\mathbi{x}_2,\cdots\mathbi{x}_2}_{\text{total $n-m$}}) \, .
\end{equation}
Since $\xi_R^{(n)}\ll1$ on large scales, we may expand (\ref{2pointcorr}) keeping up to 4-point correlation function,  $n=4$. Then, we have 
\begin{align}\label{2pointcorr2}
\xi_h(\mathbi{x}_1,\mathbi{x}_2) \approx & \left( \frac{\nu}{\sigma_R} \right)^2 \xi_R^{(2)}(r) + \left( \frac{\nu}{\sigma_R} \right)^3 \xi_R^{(3)}(\mathbi{x}_1,\mathbi{x}_1,\mathbi{x}_2)
\nonumber\\
& + \left( \frac{\nu}{\sigma_R} \right)^4 \left\{ \frac{1}{2} \left[ \xi_R^{(2)}(r) \right]^2 + \frac{1}{3}\xi_R^{(4)}(\mathbi{x}_1,\mathbi{x}_1,\mathbi{x}_1,\mathbi{x}_2) \right.
\nonumber\\
& \left. \hspace{1.7cm} + \frac{1}{4} \xi_R^{(4)}(\mathbi{x}_1,\mathbi{x}_1,\mathbi{x}_2,\mathbi{x}_2) \right\} \, .
\end{align}
We can find straightforwardly the Fourier transform of $\xi_h(\mathbi{x}_1,\mathbi{x}_2)$ and write the power spectrum of haloes as
\begin{align}\label{MLBhaloP}
P_h(k) = & b_{L}^2 P_R(k) + b_{L}^3 \int \frac{d^3q}{(2\pi)^3} B_R(\mathbi{q},-\mathbi{k},\mathbi{k}-\mathbi{q})
\nonumber\\
& + \frac{b_{L}^4}{2} \int \frac{d^3q}{(2\pi)^3} P_R(q)P_R(|\mathbi{k}-\mathbi{q}|)
\nonumber\\
& + \frac{b_{L}^4}{3} \int \frac{d^3q_1d^3q_2}{(2\pi)^3} T_R(\mathbi{q}_1,\mathbi{q}_2,-\mathbi{k},\mathbi{k}-\mathbi{q}_1-\mathbi{q}_2) \nonumber\\
&
 + \frac{b_{L}^4}{4} \int \frac{d^3q_1d^3q_2}{(2\pi)^3} T_R(\mathbi{q}_1,\mathbi{q}_2,\mathbi{k}-\mathbi{q}_1,-\mathbi{k}-\mathbi{q}_2) \, ,
\end{align}
where we have defined the linear Lagrangian bias as
\begin{equation}
b_{L} \equiv \frac{\nu}{\sigma_R} = \frac{\delta_c}{\sigma_R^2} \, .
\end{equation}

Before we proceed, let us make some comments on the third term of (\ref{MLBhaloP}). It is very well known~\citep{2006PhRvD..74j3512M} that this term gives problems on both large and small scales. On small scales the initial power spectrum is modified~\citep{2006PhRvD..73f3519C,2006PhRvD..73f3520C,2007PhRvD..75d3514M} and the integral diverges, which would not happen for the original $P_R(k)$ [see e.g. \citet{2011ApJ...727...22J}]. For this divergence in principle we should employ the full non-linear treatment of the power spectrum on small scales to which the result is very sensitive. The proper study of fully non-linear regime on small scales is beyond the scope of the present article and we will not discuss it any further. On large scales, we can cope with the constancy of this term by ``renormalization'', i.e. absorbing it into the shot-noise term and subtract this contribution by replacing $P_R(|\mathbi{k}-\mathbi{q}|)$ with $P_R(|\mathbi{k}-\mathbi{q}|) - P_R(q)$. Then we recover the linear theory as $k\to0$. This is usually the case when e.g. one studies the non-linear bias for Gaussian density field~\citep{2009ApJ...691..569J}. In the following, for simplicity, we neglect this contribution to the bias. As we will see shortly, we can find an interesting contribution of $\fnl$ and $\tnl$ to bias on large scales where the term we are going to neglect does not play a significant role after renormalization.

\section{Scale dependent bias}
\label{sec:dependentbias}

With the halo power spectrum (\ref{MLBhaloP}), we can find the contributions from primordial non-Gaussianity by using (\ref{bihalo}) and (\ref{trihalo}). In a more convenient form to read the bias, with the redshift factor explicit, we can find
\begin{align}\label{Lbias}
P_h(k) = & b_{L}^2(z) P_R(k,z) \left [ 1  + 4\fnl\frac{\delta_c\calF_R(k)}{D(z)\calM_R(k)} \right.
\nonumber\\
& \left. + 6\gnl\frac{\delta_c^2\calG_R(k)}{D^2(z)\calM_R^2(k)} + \frac{25}{9}\tnl\frac{\delta_c^2\calT_R(k)}{D^2(z)\calM_R^2(k)} \right] \, ,
\end{align}
where $D(z)$ is the linear growth function and $b_{L}(z)$ is the linear Lagrangian bias dependent on the redshift, which is given by
\begin{equation}
b_{L}(z) = \frac{\delta_c}{D^2(z)\sigma_R^2} \, .
\end{equation}
Here, the form factors are given by
\begin{align}
\label{formFR}
\calF_R(k) = & \frac{1}{\sigma_R^2} \int \frac{d^3q}{(2\pi)^3} \calM_R(q)\calM_R(|\mathbi{k}-\mathbi{q}|)
\nonumber\\
& \times P_\phi(q) \left[ 1 + \frac{P_\phi(|\mathbi{k}-\mathbi{q}|)}{2P_\phi(k)} \right] \, ,
\\
\label{formGR}
\calG_R(k) = & \frac{1}{\sigma_R^4} \int \frac{d^3q_1d^3q_2}{(2\pi)^{3\cdot2}} \calM_R(q_1)\calM_R(q_2)P_\phi(q_1)P_\phi(q_2)
\nonumber\\
& \times \left\{ \calM_R(k)\calM_R(|\mathbi{k}-\mathbi{q}_{12}|) \left[ 1 + \frac{P_\phi(|\mathbi{k}-\mathbi{q}_{12}|)}{3P_\phi(k)} \right] \right.
\nonumber\\
& \hspace{0.5cm} + \calM_R(|\mathbi{k}-\mathbi{q}_1|)\calM_R(|\mathbi{k}+\mathbi{q}_2|) \frac{P_\phi(|\mathbi{k}-\mathbi{q}_1|)}{4P_\phi(k)} 
\nonumber\\
& \hspace{0.8cm} \times \left.
\left[ 1 + \frac{P_\phi(|\mathbi{k}+\mathbi{q}_2|)}{P_\phi(q_2)} + \frac{2P_\phi(|\mathbi{k}+\mathbi{q}_2|)}{P_\phi(|\mathbi{k}-\mathbi{q}_1|)} \right] \right\} \, ,
\\
\label{formTR}
\calT_R(k) = & \frac{1}{2\sigma_R^4} \int \frac{d^3q_1d^3q_2}{(2\pi)^{3\cdot2}} \calM_R(q_1)\calM_R(q_2)P_\phi(q_1)P_\phi(q_2)
\nonumber\\
& \times \left\{ 4\calM_R(k)\calM_R(|\mathbi{k}-\mathbi{q}_{12}|) \frac{P_\phi(|\mathbi{k}-\mathbi{q}_1|)}{P_\phi(k)} \right.
\nonumber\\
& \hspace{0.5cm}
\times
 \left[ 1 + \frac{P_\phi(k)P_\phi(|\mathbi{k}-\mathbi{q}_{12}|)}{P_\phi(q_1)P_\phi(q_2)} \right]
\nonumber\\
&  + \calM_R(|\mathbi{k}-\mathbi{q}_1|)\calM_R(|\mathbi{k}+\mathbi{q}_2|) 
\nonumber\\
& \hspace{0.5cm}
\times
\left[ 1 + \frac{P_\phi(|\mathbi{k}+\mathbi{q}_2|)}{P_\phi(q_2)} + \frac{P_\phi(|\mathbi{k}-\mathbi{q}_1+\mathbi{q}_2|)}{P_\phi(k)} \right.
\nonumber\\
&\hspace{1cm}
+ \frac{P_\phi(q_{12})P_\phi(|\mathbi{k}-\mathbi{q}_1|)}{P_\phi(k)P_\phi(q_2)} 
\nonumber\\
&\hspace{1cm} \left.\left.
+ \frac{P_\phi(q_{12})P_\phi(|\mathbi{k}+\mathbi{q}_2|)}{P_\phi(k)} \frac{P_\phi(q_1)+P_\phi(q_2)}{P_\phi(q_1)P_\phi(q_2)} \right] \right\} \, .
\end{align}
Now, let us relate the Lagrangian bias $\mathfrak{b}_L^2 = P_h(k,z)/P_R(k,z)$ to the Eulerian bias $b_E$ as $b_E \equiv 1 + \mathfrak{b}_L$. This is based on the standard assumptions that haloes and the underlying dark matter move in the same way. Then, with the Taylor expansion of (\ref{Lbias}), we can find, with $b_0 \equiv 1 + b_L$,
\begin{align}\label{Ebias}
b_E = & b_0 + \Delta b \, ,
\\
\Delta b \approx &
2\fnl\frac{(b_0-1)\calF_R(k)\delta_c}{D(z)\calM_R(k)} + 3\gnl\frac{(b_0-1)\calG_R(k)\delta_c^2}{D^2(z)\calM_R^2(k)}
\nonumber\\
&
+ \left[ \frac{25}{18}\tnl\calT_R(k) - 2\fnl^2\calF_R^2(k) \right] \frac{(b_0-1)\delta_c^2}{D^2(z)\calM_R^2(k)} \, .
\end{align}

\begin{figure}
  \begin{center}
    \includegraphics[width=80mm]{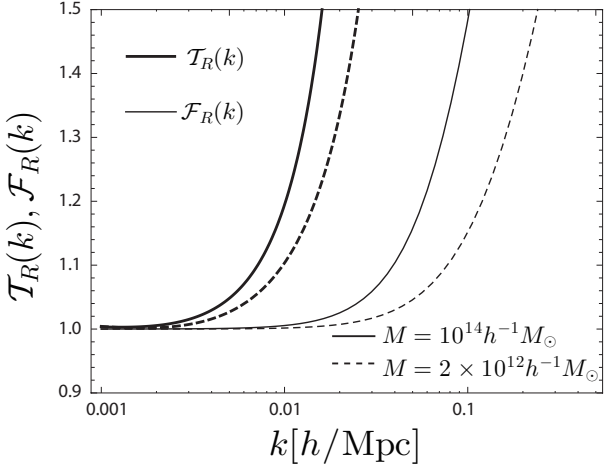}
  \end{center}
  \caption{Form factors, $\calF_R(k)$ and $\calT_R (k)$, are plotted with changing the smoothing scale:
  $M=10^{14} h^{-1} M_{\odot}$ and $2 \times 10^{12} h^{-1} M_{\odot}$. We have used the top-hat window function.
  }
  \label{fig:F_T}
\end{figure}

\begin{figure}
   \begin{center}
     \includegraphics[width=80mm]{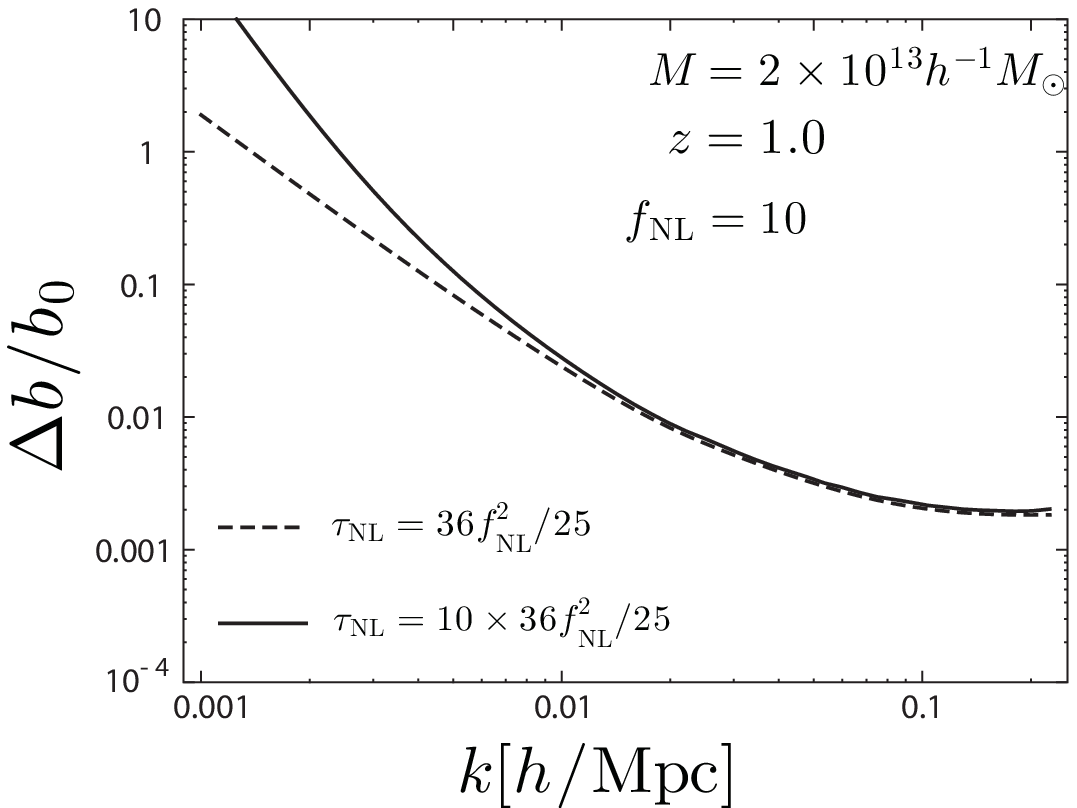}
   \end{center}
   \caption{Scale dependent bias for $\tnl=10(6\fnl/5)^2$ (solid line) and $\tnl=(6\fnl/5)^2$ (dashed line) at $z=1.0$ for $\fnl=10$ 
   and $M = 2 \times 10^{13}h^{-1} M_\odot$.}
   \label{fig:sdtNL}
\end{figure}

\begin{figure}
   \begin{center}
     \includegraphics[width=80mm]{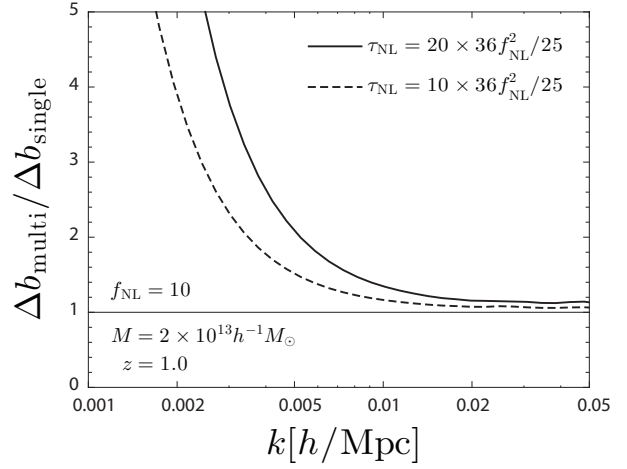}
   \end{center}
   \caption{The ratio between $\Delta b$ in the multiple source case and that in the single source case at $z=1.0$ for $\fnl=10$ 
   and $M = 2 \times 10^{13}h^{-1} M_\odot$. The solid line is for $\tnl = 20 (6\fnl/5)^2$ and the dashed line for $\tnl = 10 (6\fnl/5)^2$.}
   \label{fig:ratio}
\end{figure}

Let us focus on the last term in the right hand side of the above equation which depends on the non-linearity parameters $\fnl$ and $\tnl$.
The form factor $\calF_R(k)$ given by (\ref{formFR}) is well known to approach 1 on large scales. Another form factor $\calT_R(k)$ we have presented in (\ref{formTR}) becomes close to 1 on large scales, $k\lesssim0.01h\mathrm{Mpc}^{-1}$. $\calT_R(k)$ is plotted in Figure~\ref{fig:F_T} as a function of $k$, in comparison with $\calF_R(k)$.
Hence, this term is approximately proportional to $\tnl - (6\fnl/5)^2$ on large scales.
Thus, depending on whether the relation $\tnl = (6\fnl/5)^2$ holds or not,
we may have additional contribution to the scale dependent bias. This suggests a new possibility of probing the relation between the non-linearity parameters $\fnl$ and $\tnl$, and in turn of constraining the generating mechanism of primordial non-Gaussianity.
From (\ref{ineq}), we can see that if primordial non-Gaussianity is originated from multiple sources, the bias on large scales is enhanced compared with the single source case.
In Figure~\ref{fig:sdtNL}, with $\fnl = 10$ and $M = 2 \times 10^{13}h^{-1}M_\odot$, we show the scale dependent bias for $\tnl=10(6\fnl/5)^2$ (solid line) and $\tnl=(6\fnl/5)^2$ (dashed line) at $z=1.0$.
Here, $M$ denotes the mass given by $ M = 4\pi \bar{\rho} R^3 /3$ with the background matter density $\bar{\rho}$ and the smoothing scale $R$.
From this figure, we can find that for $\tnl=10(6\fnl/5)^2$
the effect of the non-zero $\tnl$ on the scale dependent bias
becomes marked on $k \leq 0.01 h / {\rm Mpc}$: at $k = 0.005 h / {\rm Mpc}$,
$\Delta b / b_0$ is about $50 \%$ larger than that for $\tnl=(6\fnl/5)^2$.
In Figure~\ref{fig:ratio}, with  $\fnl = 10$ and $M = 2 \times 10^{13}h^{-1}M_\odot$,
we show the ratio between $\Delta b$ in the multiple source case and that in the single source case, $\Delta b_{\rm multi} / \Delta b_{\rm single}$.
The solid line is for $\tnl = 20 (6\fnl/5)^2$ and the dashed line for $\tnl = 10 (6\fnl/5)^2$.

\section{Conclusions}
\label{sec:conclusion}

In this article, we have studied the effects of generic non-zero trispectrum of the primordial curvature perturbation on LSS, in particular, the bias of the haloes. By making use of the functional integration approach for the correlation functions of the density peaks with a high threshold at large separations, we have developed an analytic expression of the halo bias which includes scale dependence due to the local type primordial non-Gaussianity. In the presence of generic trispectrum which is parametrized by $\gnl$ and $\tnl$, we have found new scale dependent terms contributing to the bias which are unknown before. Especially, we have found that $\tnl$ appears in a combination with $\fnl$ as $\tnl^2-(6\fnl/5)^2$ on large enough scales $k\lesssim0.01h/\mathrm{Mpc}$. As shown in Figure~\ref{fig:sdtNL}, on such scales depending on the relation between $\fnl$ and $\tnl$ we have found distinguishable behaviour of bias. This suggests that the halo bias on large scales provides a very interesting and new possibility of probing the relation between $\fnl$ and $\tnl$. This relation depends on the generating mechanism of primordial non-Gaussianity, thus constraining this relation by halo bias would provide a powerful tool for studying the origin of non-Gaussianity and models of inflation. Future surveys will provide a large sample galaxies, and our finding will be useful for constraining primordial non-Gaussianity.

\subsection*{Acknowledgements}

We thank Donghui Jeong, Takahiko Matsubara, Ravi Sheth, Naoshi Sugiyama,
Takahiro Tanaka and Yoshitaka Takeuchi for useful discussion.
We acknowledge the workshop ``Cosmological Perturbation and Cosmic Microwave Background"
(YITP-T-10-05) at the Yukawa Institute for Theoretical Physics, Kyoto University, where this work was initiated.
This work was supported in part by
a Korean-CERN fellowship,
the Grant-in-Aid for Scientific research from the Ministry of Education,
Science, Sports and Culture (MEXT), Japan, No. 22340056 and
the Grant-in-Aid for the Global COE Program ``Quest for Fundamental Principles in the
Universe: from Particles to the Solar System and the Cosmos'' from
MEXT, Japan.

\vspace{1cm}\noindent CERN-PH-TH/2011-145

\label{lastpage}

\end{document}